# Stochastic Gene Expression as a Many Body Problem

Masaki Sasai[*†] and Peter G. Wolynes[†]

*Graduate School of Human Informatics, Nagoya University, Nagoya 464-8601, Japan
†Department of Chemistry and Biochemistry and Center for Theoretical Biological Physics, University of California at San Diego, La Jolla, CA 92093, USA

**Abstract**

Gene expression has a stochastic component owing to the single molecule nature of the gene and the small number of copies of individual DNA binding proteins in the cell. We show how the statistics of such systems can be mapped on to quantum many body problems. The dynamics of a single gene switch resembles the spin boson model of a two site polaron or an electron transfer reaction. Networks of switches can be approximately described as quantum spin systems by using an appropriate variational principle. In this way the concept of frustration for magnetic systems can be taken over into gene networks. The landscape of stable attractors depends on the degree and style of frustration, much as for neural networks. We show the number of attractors, which may represent cell types, is much smaller for appropriately designed, weakly frustrated stochastic networks than for randomly connected networks.

The complexity of a cell's genome is expressed through the interactions of many genes with a large variety of proteins. Understanding gene expression is, therefore, a many body problem. But what kind of many body problem? Should we think of gene expression using the metaphors and techniques of deterministic many body problems like those developed for the "clockwork Universe' of 19<sup>th</sup> century celestial mechanics? Or is it appropriate to use statistical ideas like those that form the language of condensed matter physics and physical chemistry (1)?

The deterministic view has much to recommend it. The miracles of development require intricacy and precision (2). Cell cycles, a prominent dynamic sign of life not found in inanimate matter, are often described as clocks. With the great information content of the genome now so apparent in the "postgenomic era", it is hard to resist making analogies between cells and those man made information processors, electronic computers, which grind through their programs with a determination that Laplace would have found thrilling. The stochastic view is not without merit, however. Since a gene is a molecule, the statistical fluctuations of atomism cannot be avoided, as Delbrück realized so long ago (3). The technological capabilities of modern experimental biophysics have also made the presence of stochastic behavior in cells undeniable as an experimental fact (4). Under some circumstances, the game theoretic advantage of unpredictable behavior in the predator-prey relations among single cell organisms will be clear incentives for stochasticity to have evolved adaptively. Furthermore, even when modern cells have well orchestrated patterns of gene expression, we need to know how this elegant patterning can have been achieved in the light of there being both specific and non-specific interactions of DNA binding proteins with the myriad possible similar but nevertheless incorrect sites along the genome, many of which remain silent.

The main purpose of this paper is to begin the exploration of stochastic gene expression by developing an analogy to quantum many body problems. Theoretical work on stochastic models of gene expression has been dominated by simulation approaches (5-8). Owing to the intricate connectivity of real gene networks, computer simulations are doubtless necessary. Yet, by themselves they do not provide an easy route to visualizing the basic emergent principles at work. Analytical approaches to stochastic gene systems have focused on single switches using methods that are not easy to generalize to the description of a complete network (9,10). In this paper we will see how the discrete nature of the binding sites on the DNA and the finite small numbers of transcription factor proteins in a cell can be easily accommodated using a Master equation containing operators like those encountered in describing quantum many particle systems. The dynamics of the DNA binding sites (genes) will be described by quantum spin operators while the fluctuations in protein concentrations in the cell will be described using creation/annihilation operators analogous to those for bosonic harmonic oscillators. The analogy between discrete number fluctuations in chemical kinetics and quantum mechanics has been uncovered many times (11-13) and is reviewed by Mattis and Glasser (14). For gene expression problems in particular, this analogy provides an immediate connection to well-studied many body problems. A single genetic switch becomes equivalent to the spin-boson problem that features in the theory of polarons in solid state physics (15) and electron transfer in chemistry (16). Bare switches are dressed by a "proteomic atmosphere" much as electrons in insulating solids are accompanied by a cloud of phonons. In the quantum analogy switches interact through the virtual emission and reabsorption of protein fluctuations. Using this formalism, a multi-switch network can be described using the language of magnets and spin models of neural networks. Finally, through this analogy the steady states of a stochastic genetic network can be, in some approximations at least, described in landscape terms using precise mathematics rather than in the metaphorical way that has already achieved a certain popularity.

Beyond its linguistic advantages, this analogy allows the well-developed approximation methods used for quantum many body problems to be brought to bear on the gene expression problem. These methods can be based on path integrals (17), re-summed diagrammatic perturbation theory (18) and variational schemes (19). In using these tools, one significant difference from ordinary quantum many body problems must be noted, however: the effective Hamiltonian for the Master equation is not Hermitian. This reflects the far-from equilibrium nature of these systems. Owing to non Hermiticity, the mathematical formulation of the traditional approximations must be re-examined. In practice, also the quality of approximations well established in quantum many body science will have to be re-evaluated. In addition, for gene expression a wider range of phenomena occurs than comes up in traditional solid state physics, e.g. steadily oscillating states. Yet much insight from quantum many body theory can be brought over intact.

We illustrate the utility of this approach by providing a fresh perspective on a central problem of cell biology – the stability of cell types. Multicellular, eukaryotic organisms contain a relatively modest number of cell types by which we mean groups of cells that have a sensibly common pattern of which proteins are actually expressed. The small number of different cell types is puzzling. Just as in the famed Levinthal paradox of protein folding, if each gene switch, of which there must be hundreds, can be in two states, either "on" or "off", why aren't there of order $2^{100}$ cell types? This huge number would be possible if the switches were deterministic and noninteracting. This issue has been raised for interacting deterministic switches using the Kauffman *N-K* model (20). For this model an unusual fine tuning of parameters seems to be needed to obtain only a moderate number of expression patterns (21), although this may be avoided by invoking a scale-free topology of the network (22). Another possibility is that the number of potential cell types actually realized reflects the determinism of the developmental program found in the embryo. Of course this determinism raises further issues of the stochastic stability of a sequence of events rather than of a steady state. Here we explore a third possibility. We exhibit a family of models of stochastic gene expression for which the cell-type number paradox can be resolved using the concept of frustration for the multi-spin system corresponding to the gene network. As for the folding (23) problem the capacity to address a central paradox of gene expression in quantitative terms promises a starting point for practical problems of characterizing the class of genetic networks that actually describe.

First, we describe the many body analogy for a single switch with a proteomic atmosphere emphasizing the connections with the spin-boson problem. We also describe switch interactions in this framework. Following an approach of Eyink (24), we then formulate a variational approximation for the non-hermitian many body problem of a single switch and discuss an analogy to the Hartree approximation for many interacting switches. A landscape description arises naturally in this approximation. Then we explore the phase diagram of a single switch. We propose a very simple network topology where the Hartree approximation should be valid and characterize the phase diagram highlighting the range of parameters where the cell type paradox is resolved. Finally, we discuss the prospects for using these ideas to provide a general landscape picture for gene expression, for quantifying the response and fluctuation of such networks, and for computing their long term stability.

## Spin-Boson Formalism for Stochastic Switches

A variety of mechanisms for individual gene switches have been elucidated by molecular biologists. These involve the DNA directed synthesis of proteins that themselves bind to the DNA thereby turning up or down the synthesis rate. Protein synthesis itself is not simple because of the intrinsic time delays of serial synthesis and the intervening step of synthesizing messenger RNA. Since our goal here is to illustrate the mathematical tools, we will ignore these doubtless important complications. They can be included simply by introducing more species. We will describe only the simplest switches here and concentrate on a switch architecture with symmetry properties that makes our later discussion of networks more transparent. This exposition should enable the reader to see how to write the equations for any known biochemical model for a single switch or network.

The most complete description of a simple stochastic gene switch would be a path probability describing the joint probabilities at various times of the DNA operator sites being occupied by ligand proteins and for those same times the numbers of the different ligand molecules in the cell. We assume the binding proteins are well-mixed in the cell. We can extend the formalism to a completely field theoretical description of protein concentrations at different points in the cell, if needed to account for incomplete mixing. If we ignore time delays, the Markovian nature of process ensures that the path probabilitiy can be obtained from an operator description of the Master equation that describes at any one time the joint probability of the DNA occupation and the protein numbers. The two valued-ness of the DNA occupation at a site makes it convenient to describe this probability as a spinor quantum state. For concreteness consider first a single binding site gene switch. The DNA site has two states, $S=1$ active, without a bound repressor and $S=0$, inactive, with repressor bound. The 2 component vector $\mathbf{P}(n,t) = (P_1(n,t), P_0(n,t))$ expresses the joint probability of the DNA binding state and the number $n$ of proteins in the cell. The rate of protein synthesis $g_S$ depends on the DNA state, the degradation rate $k$ is independent of $S$. When $S$ is fixed, then the master equation describes a simple birth-death process (for each value of $S$) and can be formulated as a difference equation. Consider the case that the product of the gene is a repressor protein which binds to the operator site to change its own activity. Then binding of protein occurs with a rate dependent on *n*, *h(n)*, while unbinding has a rate *f*. These latter processes transfer probability between the two components of the state vector. The master equation can be written as

$$\frac{\partial}{\partial t}\mathbf{P}(n,t) = \begin{pmatrix} g_1 & 0 \\ 0 & g_0 \end{pmatrix}\{\mathbf{P}(n-1,t) - \mathbf{P}(n,t)\} \\ + k\{(n+1)\mathbf{P}(n+1,t) - n\mathbf{P}(n,t)\} + \begin{pmatrix} -h(n) & f \\ h(n) & -f \end{pmatrix}\mathbf{P}(n,t)$$
(1)

The analogy to quantum system is most apparent when we express the difference operations using operators for the creation and annihilation of protein molecules. Such a notation was introduced by Doi (11) and used extensively by Zeldovich and coworkers (12,13) to describe the small number asymptotics of diffusion limited reactions. We follow the notation and conventions outlined in (14). These differ somewhat from those ordinarily used in quantum mechanics. For each protein concentration a creation and an annihilation operator are introduced such that $a^\dagger | n \rangle = | n+1 \rangle$ and $a | n \rangle = n | n-1 \rangle$. These satisfy $[a, a^\dagger]=1$. For a process only involving a single protein particle number, the state vector is $\psi = \Sigma_n P(n,t) | n \rangle$ where *P(n,t)* is the probability of having precisely *n* particles. The master equation (1) is written $\partial \psi / \partial t = \Omega \psi$ using a spinor hamiltonian for the dynamics of the DNA coupled to the proteins. $\Omega$ is a non Hermitian "hamiltonian" operator. $\Omega$ for this simple gene switch is

$$\Omega = (\bar{g} + \delta g \sigma_z)(a^\dagger - 1) + k(a - a^\dagger a) + \mu^+(-1 + \sigma_x) + \mu^-(-i\sigma_y - \sigma_z),$$
(2)

$$\text{where } \sigma_x = \begin{pmatrix} 0 & 1 \\ 1 & 0 \end{pmatrix}, \; i\sigma_y = \begin{pmatrix} 0 & 1 \\ -1 & 0 \end{pmatrix}, \; \sigma_z = \begin{pmatrix} 1 & 0 \\ 0 & -1 \end{pmatrix},$$

and $\bar{g} = (g_1 + g_0)/2$, $\delta g = (g_1 - g_0)/2$, $\mu^+ = (h(a^\dagger a) + f)/2$ and $\mu^- = (h(a^\dagger a) - f)/2$. In this operator formalism averages are obtained by taking the scalar product with the bra $\langle 0 | e^a$.

$\Omega$ is the Hamiltonian of a Spin-boson Hamiltonian, albeit an explicitly non Hermitian hamiltonian. Probability is conserved owing to the unusual scalar product. We see that in this representation the "spin" state of the DNA polarizes a "proteomic" atmosphere, just as in conventional electron transfer charge motion of an acceptor-donor pair distorts the surrounding lattice by coherently creating phonons. This distortion of the proteomic atmosphere acts to stabilize the two distinct states of the switch allowing the switch states to be much more stable and change much more slowly than the chemical off-rate alone might suggest. As in electron transfer the change from one switch state to another can be viewed either in an adiabatic or nonadiabatic representation depending on whether the DNA occupation variable always can follow the protein number variable. Generally the adiabatic limit is thought to apply, but this may not be the universal rule. In the study of charge transfer that different approximations that turn to be exact in one limit are still useful to describe the kinematics in the other limit.

Gene network is made up of elements containing binding sites controlling protein production. The state space of the entire network is the direct product of the state spaces for each element and the non Hermitian hamiltonian operator $\Omega$ is the sum of terms $\Omega_i$ describing each such element $\Omega = \Sigma_i \Omega_i$. Explicitly we have

$$\Omega_i = (\bar{g}_i + \delta g_i \sigma_{i,z})(a_i^\dagger - 1) + k_i(a_i - a_i^\dagger a_i) + \mu_i^+(-1 + \sigma_{i,x}) + \mu_i^-(-i\sigma_{i,y} - \sigma_{i,z}), \qquad (3)$$

with $\mu_i^\pm = \left(h_i(a_j^\dagger a_j) \pm f_i\right)/2$. $\delta g_i$ in Eq.3 is positive when the transcription factor that binds to the *i*th gene is a repressor and negative when it is an activator. In a network of many gene elements each element not only interacts with its directly generated proteomic atmosphere as in the polaron, but interactions between gene elements occur through the exchange of proteins - the quanta of the proteomic fields. It is useful to visualize this in the representation where the DNA state changes slowly. In this case, we generate indirect spin-spin interactions in the non Hermitian hamiltonian, just as in the theory of condensed phase magnets. These can be considered ferromagnetic or anti-ferromagnetic depending on whether the exchanged protein is a repressor or an activator. A nicely symmetric two element switch model is illustrated in Fig.1a. Three or more interacting components give rise to the possibility of frustrated or unfrustrated interactions in the sense of whether the corresponding spin-spin interactions lead to coherent or incoherent activation patterns. This can be decided easily by taking the product of the induced spin-spin interactions around a closed loop, a positive product being unfrustrated, a negative product tending to give multiple states or cycles.

Many gene switches involve multimers of individual proteins or several gene products. In this case the on-rates simply depend on higher polynomials of the relevant protein number operators: When a monomer of the product of the *j*th gene binds to the *i*th gene site, $h_i(n_j) = h_{ij} a_j^\dagger a_j$, and when a dimer of the *j*th product binds to the *i*th site, $h_i(n_j) = h_{ij}(a_j^\dagger a_j)^2$.

**Variational Approach in Spin-Boson Formalism**

A variational method developed by Eyink (24) in a different context provides a particularly lucid set of approximations. The master equation is equivalent to the functional variation $\delta \Gamma / \delta \psi^L = 0$ of an effective "action" $\Gamma = \int dt <\psi^L | (\partial_t - \Omega) | \psi^R >$ with $\psi = |\psi^R>$. This functional variation can be reduced to a set of finite dimensional equations by representing $\psi^L$ with parameters, $\alpha^L=(\alpha_1^L, \alpha_2^L, ..., \alpha_K^L)$, as $\psi^L(\alpha^L)$ and $\psi^R$ as $\psi^R(\alpha^R)$. Here, $\psi^L(\alpha^L=0)$ is set to be consistent with the probabilistic interpretation $<\psi^L(\alpha^L=0)| \psi^R(\alpha^R)>=1$. In the spin-boson formalism this constraint implies $<\psi^L(\alpha^L=0)| = <0|\exp(\Sigma_i a_i)$. The condition that this physically sensible $\psi^L$ is an extremum of the action is

$$\left( \sum_{l=1}^{K} <\frac{\partial \psi^L}{\partial \alpha_m^L} | \frac{\partial \psi^R}{\partial \alpha_l^R}> \frac{d\alpha_l^R}{dt} - <\frac{\partial \psi^L}{\partial \alpha_m^L} | \Omega | \psi^R > \right)_{\alpha_m^L=0} = 0, \quad \text{for} \quad m=1,...K. \quad (4)$$

To apply Equation 4, explicit functional forms of $\psi^L(\alpha^L)$ and $\psi^R(\alpha^R)$ have to be given. In the simple birth-death problem, for example, the probability distribution $P(n,t)$ to find $n$ particles should be Poisson at large $t$, $P(n)=(X^n/n!) \exp(-X)$ with a mean $X$, so that the state vector $\psi=\Sigma_n P(n,t)|n>$ approaches a "coherent state", $\psi=\exp(X(a^\dagger-1))|0>$. To analyze the relaxation toward this stationary state, one may choose the functional form $\psi^R=\exp(X(t)(a^\dagger-1))|0>$, with $\alpha^R=X(t)$. The corresponding $\psi^L$ is chosen so as to make the variation equations simple: a reasonable choice is $\psi^L=<0|\exp(a)\exp(\lambda a)$ with $\alpha^L=\lambda$. This "coherent state Ansatz" for $\psi^R$ and $\psi^L$ can be taken further to describe more complex processes using for example "squeezed states".

With the coherent state Ansatz for the single gene problem, the state vector has two components corresponding to $S=1$ and 0.

$$|\psi^R> = \begin{pmatrix} C_1 \exp(X_1(a^\dagger-1))|0> \\ C_0 \exp(X_0(a^\dagger-1))|0> \end{pmatrix},$$

$$<\psi^L| = \begin{pmatrix} <0|e^a \exp(\alpha_1 + \lambda_1 a) & <0|e^a \exp(\alpha_0 + \lambda_0 a) \end{pmatrix}. \quad (5)$$

where, $C_1$ and $C_0$ are the probabilities of the two DNA binding states $S=1$ and 0, respectively. With this ansatz, the coupled dynamics of the DNA binding state and the protein distribution is described as the motion of wavepackets with amplitudes $C_1(t)$ and $C_0(t)$ and means at $X_1(t)$ and $X_0(t)$.

A straightforward choice of the trial state vector for a gene network is a Hartree-type product of single spin-boson vectors:

$$|\psi^R> = \prod_i |\psi^R(i)>, \quad <\psi^L| = \prod_i <\psi^L(i)|, \tag{6}$$

where $|\psi^R(i)>$ and $<\psi^L(i)|$ are vectors of the $i$th element. With the coherent state Ansatz, $|\psi^R(i)>$ and $<\psi^L(i)|$ have the same form as in Eq.5 with replacement of $a^\dagger$ with $a_i^\dagger$ and $C_1$ with $C_1(i)$, and so on. A time dependent Hartree approximation of the network is obtained by putting Eqs.3 and 6 into Eq.4. In the Hartree approximation the $i$ and $j$th elements interact with each other through terms like $h_j(a_i^\dagger a_i)$ in $<\psi^L(i)\psi^L(j)|\Omega|\psi^R(i)\psi^R(j)>$. In this way genes couple through the proteomic atmosphere which is here represented by field operators.

By introducing an effective "potential energy", the term $<\partial\psi^L/\partial\alpha_m^L|\Omega|\psi^R>_{\alpha_m^L=0}$ in the Hartree equation can be expressed by a sum of derivatives of a potential energy for each switch and a residual term. Regarding the residual term as a noise, it is natural to use the energy landscape language to describe behaviors of the network. When interactions among gene elements are unfrustrated, one may expect that the landscape of the effective potential energy is dominated by a small number of distinct valleys. When the network involves a sufficient number of frustrated loops, on the other hand, the landscape should be rugged and the time dependent Hartree trajectory would be trapped into one of many local minima or would never settle into a stationary state owing to the lack of detailed balance.

## Phase diagram of a Single Switch

The gene circuit shown in Fig.1a is composed of two interacting genes. The product of gene A is a repressor that binds to gene B and the product of gene B is a repressor that binds to gene A. This circuit was experimentally implemented in *E. coli* plasmids and shown to work as a toggle switch (25). In one state, gene A is more active than gene B and in the other state gene B is more active. An inducer which changes the activity of a repressor can toggle between two states. Stochastic fluctuations in switching were numerically simulated in the adiabatic regime (26). Here we apply the Hartree approximation to this circuit. The derived phase diagram illustrates how the adiabaticity affects the switching behavior.

The two genes are assumed to be symmetrical with the same production rates, $\bar{g}(A) = \bar{g}(B) = \bar{g}$ and $\delta g(A) = \delta g(B) = \delta g$, and the same unbinding-rates, $f_A = f_B = f$. The degradation rates of two proteins are also assumed to be same, $k_A = k_B = k$. We consider the case that repressors bind to DNA in a dimer form with the binding rates $h_A = h(a_B^\dagger a_B)^2$ and $h_B = h(a_A^\dagger a_A)^2$. Using scaled parameters makes the description of the results more transparent $\omega = f/k$, $X^{eq} = f/h$, $X^{ad} = \bar{g}/k$, and $\delta X = \delta g/k$. $\omega$ is an "adiabaticity" parameter representing the relative speed of the DNA state alterations to the rate of the protein number fluctuations. $X^{eq}$ measures the tendency that proteins are unbound from DNA.

Using Eqs.3,4, and 6, the Hartree equations are derived. In the adiabatic limit of small $\omega$, they have a stationary solution with $X_1(A) = X_1(B) = X^{ad} + \delta X$ and $X_0(A) = X_0(B) = X^{ad} - \delta X$. The order parameter of the switching ability, $\Delta C = C_1(A) - C_1(B)$, is $\Delta C = 0$ in this case. Effects of the small protein number become more evident when $X^{ad}$ is smaller. In the limit of small $X^{ad}$, fluctuations are large, which also leads to $\Delta C = 0$.

The Hartree equation for $\Delta C$ can be written as $(X^{eq}/\omega)(d\Delta C/dt) = -\partial V / \partial \Delta C$ + (residual terms). The shape of the effective potential energy $V$ is shown in Fig.2a as a function of $\Delta C$. The stationary solution of the Hartree equations corresponds to a minimum of this potential energy. The potential energy has a single minimum in the regime of small $\omega$ or $X^{ad}$, but has double minima when $\omega$ and $X^{ad}$ are large. By numerically solving the stationary Hartree equations, $|\Delta C|$ is plotted on the $\omega$ - $X^{ad}$ plane in Fig.2b. When $\omega$ or $X^{ad}$ is small, the circuit fluctuates with equal probability between two states and does not acts as a stable switch. At the phase boundary of Fig.2b, the $\Delta C = 0$ state becomes unstable and bifurcates into two equivalent states with positive and negative $\Delta C$. In the nonzero $\Delta C$ phase the circuit takes either of two states and works as a toggle switch between them.

## Switch Interaction, Network Topology, and the Attractor Landscape

With the Hartree approximation it is possible to analyze a large scale network composed of many genes. Here, for simplicity, we consider networks whose elements are the toggle switches discussed in the last section. When there are $N$ uncoupled switches, the network potentially has $2^N$ states This is a huge number even for a moderate $N$. We will see this huge number may be much reduced when the network of switches interacts in only a weakly frustrated manner.

The simplest design for the network is shown in Fig.1b. The $i$th switch has two operons $A_i$ and $B_i$. $A_i$ produces both the intra- and inter-switch repressors and $B_i$ produces only the intra-switch repressor. Their production rates are controlled by the binding state of the operator site of each operon. The binding rates of repressors $h_{Ai}$ and $h_{Bi}$ at the operator sites $A_i$ and $B_i$ are

$$h_{Ai} = h(a_{Bi}^\dagger a_{Bi})^2 \quad \text{and} \quad h_{Bi} = h(a_{Ai}^\dagger a_{Ai})^2 + \frac{h'}{N}\sum_j a_{Aj}^\dagger a_{Aj}, \tag{7}$$

respectively. In $h_{Bi}$ the binding-rate of the inter-switch protein is scaled as $h'/N$ because each inter-switch repressor diffuses over $N$ switches. The Hartree equations for this network have a stationary ferromagnetic solution with $\Delta C = C_1(A_i) - C_1(B_i) > 0$ for all $i$ when $\omega$ and $X^{ad}$ are large enough. The phase diagram on the $\omega$-$X^{ad}$ plane shows that the region of the ferromagnetic phase with $\Delta C > 0$ is wider than the switching region $\Delta C \neq 0$ of the single switch in Fig.2b. Such ferromagnetic networks can also be designed by using activator proteins.

Denoting the switch state by $\xi_i = \text{sgn}(C_1(A_i) - C_1(B_i))$, every switch in the network of Fig.1b is homogeneous, $\xi_i = 1$, in the ferromagnetic phase. Heterogeneous switching states with any designed pattern, $\xi_i^0 = \pm 1$, are also possible when interactions are transformed from the all ferromagnetic of Fig.1b into another set of unfrustrated interactions of Fig.1c. This is analogous to the so-called Mattis ferromagnet (27). Here, repressors bind to $A_i$ and $B_i$ with the binding rates $h_{Ai}$ and $h_{Bi}$,

$$h_{Ai} = h(a_{Bi}^\dagger a_{Bi})^2 + \frac{1-\xi_i^0}{2}\frac{h'}{N}\sum_j \left( \frac{1+\xi_j^0}{2} a_{Aj}^\dagger a_{Aj} + \frac{1-\xi_j^0}{2} a_{Bj}^\dagger a_{Bj} \right),$$

$$h_{Bi} = h(a_{Ai}^{\dagger} a_{Ai})^2 + \frac{1+\xi_i^0}{2} \frac{h'}{N} \sum_j \left( \frac{1+\xi_j^0}{2} a_{Aj}^{\dagger} a_{Aj} + \frac{1-\xi_j^0}{2} a_{Bj}^{\dagger} a_{Bj} \right). \qquad (8)$$

This target-dependent transformation of interactions from Eq.7 to Eq.8 is analogous to a gauge transformation in the spin system. By making this transformation, an unfrustrated set of interactions will have a frozen state with a different arrangement of bound and unbound sites. The Hartree equations for this Mattis network have a stationary solution of $\xi_i = \xi_i^0$ in the same parameter region where the ferromagnetic solution exists.

When each operon produces multiple kinds of inter-switch repressors or activators, then the proteomic atmosphere is a superposition of many kinds of these transcription factors By having interactions which are sums of different Mattis patterns, such gene networks can exhibit multiple expression patterns analogous to the memories of a Hopfield neural network (28). Superposition of multiple interactions, however, yields both the unfrustrated and frustrated interactions. In such networks both solutions retrieving a given binding pattern and solutions which are irrelevant to that binding pattern may coexist.

To describe the phase diagram we consider a network composed of repressors only. The network may be designed to have superimposed interactions of $p$ binding patterns $\xi_i^l$ with $l=1$-$p$. We assume binding of the intra-switch repressor and the inter-switch repressor at the operator site cause the same effect on the protein production rate of each operon. The degradation rate $k$ and the unbinding-rate $f$ are assumed to be same for all proteins. Then, the relevant scaled parameters for the phase diagram are $\omega$, $X^{eq}$, $X^{ad}$, and $\delta X$ of the last section and $\Lambda^{eq} = h'/h$. The equation for $S_i = C_1(A_i) - C_1(B_i)$ is approximately derived from the Hartree equations as

$$\frac{X^{eq}}{\omega} \frac{d}{dt} S_i = -\frac{\partial}{\partial S_i} \left( -\frac{r_0}{2} S_i^2 + \frac{u}{4} S_i^4 - \frac{K}{2} \sum_{i \neq j} J_{ij} S_i S_j \right), \qquad (9)$$

where $J_{ij} = \frac{1}{N} \sum_{l=1}^{p} \xi_i^l \xi_j^l$, $K = \Lambda^{eq} \overline{C} \delta X$, $u = \delta X^2 \left\{ 1 - \left( \overline{C}/(\omega + \overline{C}) \right)^2 \right\}$, $r_0 = 4u\overline{C}^2 -$
$\left( X^- + X^{-2} + X^{eq} + p\Lambda^{eq}(X^-/2 + \delta X \overline{C}) \right)$, and $X^- = X^{ad} - \delta X$. We see that Eq.9 represents gradient descent in an energy landscape. $\overline{C} = (C_1(A_i) + C_1(B_i))/2$ is self-consistently derived from the Hartree equations.

For small $\omega$, Eq.9 has only a solution with $S_i=0$, in which elements fluctuate independently from each other. For larger $\omega$ an eigenvector of the $J_{ij}$ matrix appears that can satisfy the linearized stationary equation for $S_i$. This yields the static pattern of the switch states. Such a pattern is irrelevant to the stored binding pattern $\xi_i^l$, and should be regarded as a spin-glass solution. We expect a large number of such solutions. For general multiplicity of interactions these are exponentially many as in the Potts glass (29). Owing to the lack of the detailed balance, however, the residual noise which is neglected in Eq.9 might destabilize this solution to prevent the time-dependent Hartree trajectory from being trapped into the spin-glass pattern. The ability of the network to produce a designed pattern can be examined by introducing order parameters $m^l = \frac{1}{N}\sum_{i=1}^{N} \xi_i^l S_i$.

Generally either the $m^l$'s are all small or one of $m^l$'s is $O(1)$ and others are $O(1/\sqrt{N})$. The parameter region which allows such a dominant solution is obtained by approximately solving the self consistent equations of order parameters (30). These results are summarized in the phase diagram of Fig.3.

**Discussion**

The quantum many body analogy when combined with a variational scheme allows us to understand gene networks in landscape terms. A single switch has two attractors where the wave packet Hartree solutions correspond to an active or inactive gene. As in any mean field theory these attractors are only approximate – improbable fluctuations can take the system from one basin to the other (31,32). This dynamics will occur on a much longer time scale than the time to settle into one steady state. This activated fluctuation process is rather analogous to electron transfer kinetics and can be treated by instantons using a path integral version of the present variational approximation.

While the $N$ switch problem might have been expected to have $2^N$ attractors, the present variational formulation suggests this number will be strongly reduced if the magnetic spin problem corresponding to the network is only weakly frustrated. In this case only a small number of patterns will stably emerge on the time sale of the rapid fluctuations of a single gene switch. Again transitions between the basins of attraction found by the variational treatment can occur but on much longer time scales. It will be very interesting to see whether real gene networks have the weak frustration described here or are more nearly random (21).

Several issues in gene expression require going beyond consideration of steady attractors alone. One such issue is the escape from stable attractors already mentioned. In addition we must account for the periodic attractors that are involved in the cell cycle. Many experiments probe the response to externally provided signals that are not constant in time. Other experiments may probe endogenous fluctuations about individual attractors. Finally the most central issue is not just that of steady states but the possibility of developmental programs in which epigenetic states must follow each other in specific sequences. In all these dynamical situations it should be possible to use an analogous time dependent variational formalism to the one we used here to at least test the robustness of these temporal patterns to stochastic fluctuations.

In summary, we believe the quantum many body analogy will complement detailed stochastic modeling by providing a set of powerful mathematical tools and concepts to visualize gene expression.

## Acknowledgments

The support of NSF grants PHY0225630 and PHY0216576 to the Center for Theoretical Biological Physics is much appreciated. M. Sasai was supported by the ACT-JST program of Japan Science and Technology Corporation and grants from the Japan Society for the Promotion of Science.

**Figure Legends**

Figure 1
(**a**) The circuit of a switch composed of two symmetrical genes, each of which produces the repressor that binds to the other. (**b**) An example of the network element with "ferromagnetic" interactions. (**c**) An example of the network element with interactions of the Mattis type: Interactions depend on the target pattern of each switch, $\xi_i^0 = 1$ ($A_i$ active) or $-1$ ($B_i$ active).

Figure 2
(**a**) The effective "potential energy" $V$ for dynamics of the gene circuit Fig.1a is shown as a function of the difference $\Delta C$ in activities of two identical genes. $\omega = 0.2$. (**b**) The phase diagram of the circuit of Fig.1a. A contour map of $|\Delta C|$ is plotted on the $\omega$-$X^{ad}$ plane. When $\omega$ or $X^{ad}$ is small, the circuit shows no switching ability with $\Delta C = 0$. For larger $\omega$ and $X^{ad}$ the solution of $\Delta C = 0$ becomes unstable and bifurcates into symmetry breaking states of positive and negative $\Delta C$. The circuit works as a toggle switch between those two states. $\delta X = X^{ad}$ and $X^{eq} = 1000$ for both **a** and **b**.

Figure 3.
The phase diagram of Hopfield-type network composed of $N=200$ elementary gene switches. $p$=15 binding patterns of repressors are stored in the network. For small $\omega$ or $X^{ad}$ each element shows no switching ability. For intermediate $\omega$ and $X^{ad}$ all elements work as switches but fluctuate independently from each other. For larger $\omega$ and $X^{ad}$, the network can produce any one of $p$ designed binding patterns, *i.e.* the number of cell types is $p$. In this finite cell typenumber phase, the spin-glass solutions with random switch states coexist. $\Lambda^{eq}$=0.5, $\delta X = X^{ad}$ and $X^{eq}$=1000.

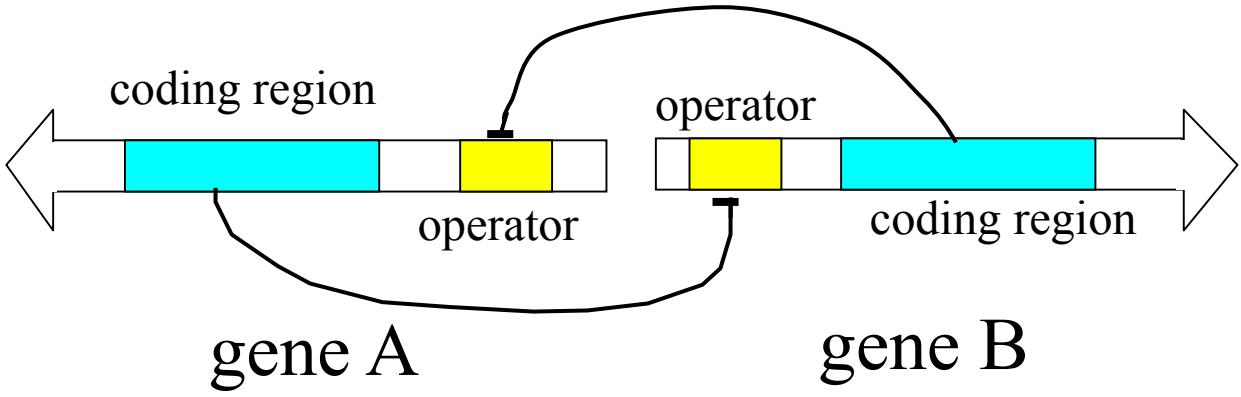

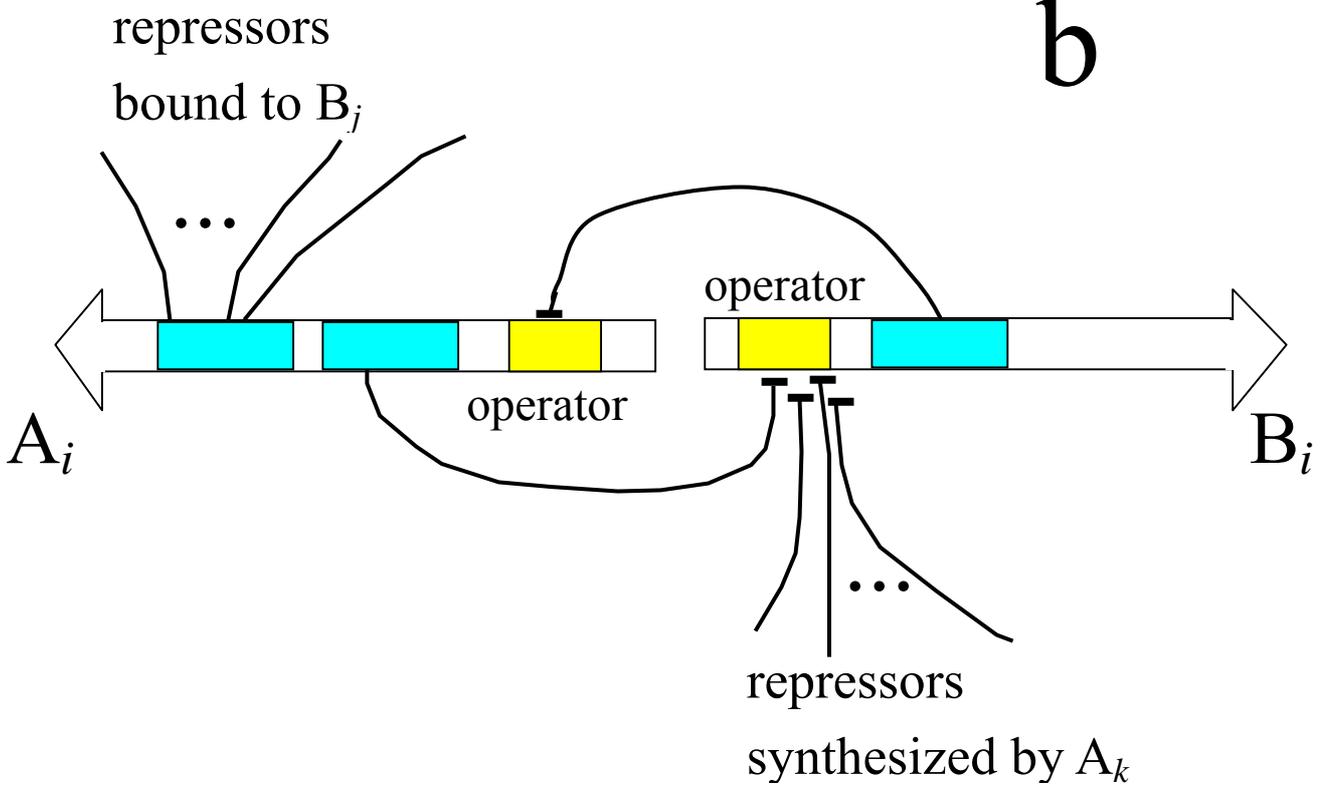

Fig.1a,b    Sasai and Wolynes

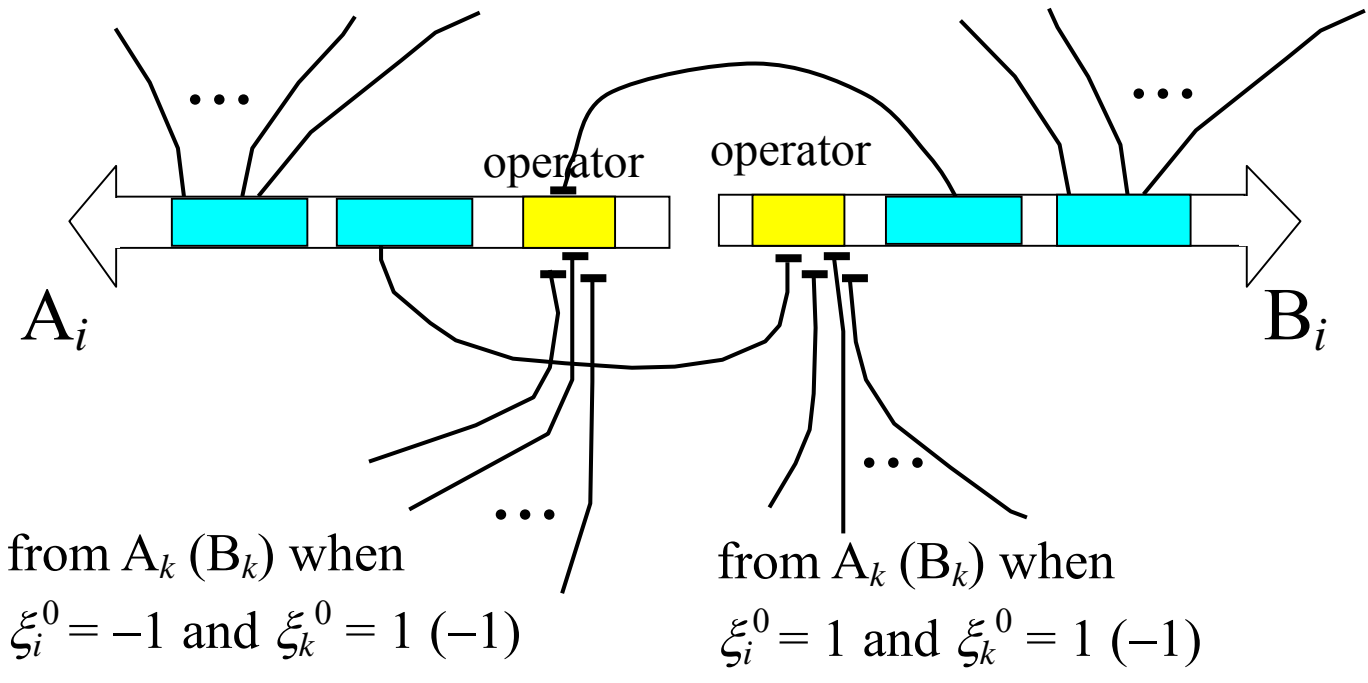

Fig.1c  Sasai and Wolynes

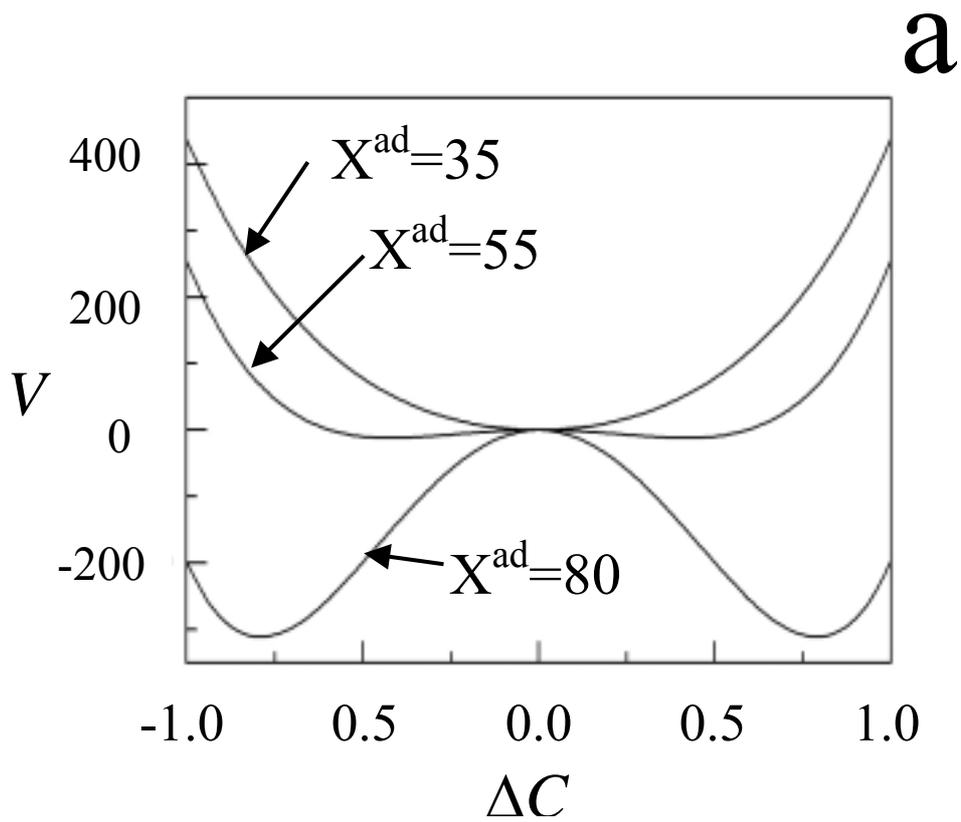

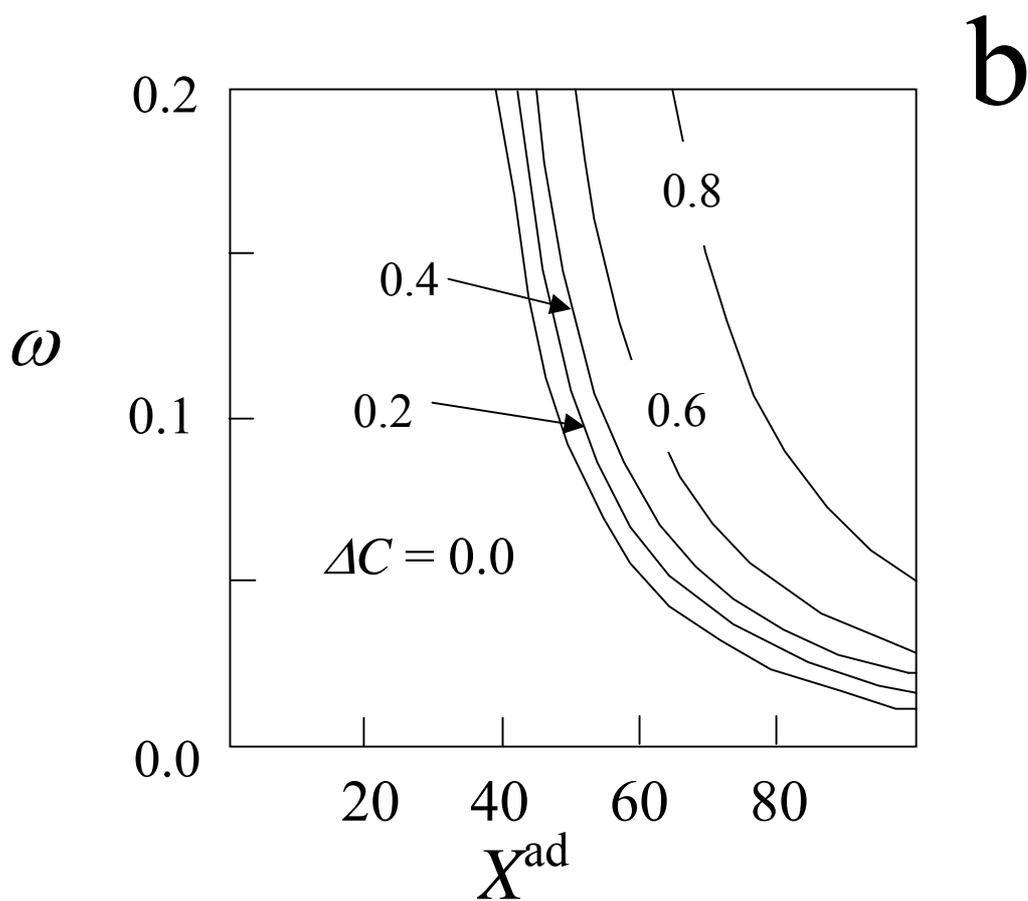

Fig.2    Sasai and Wolynes

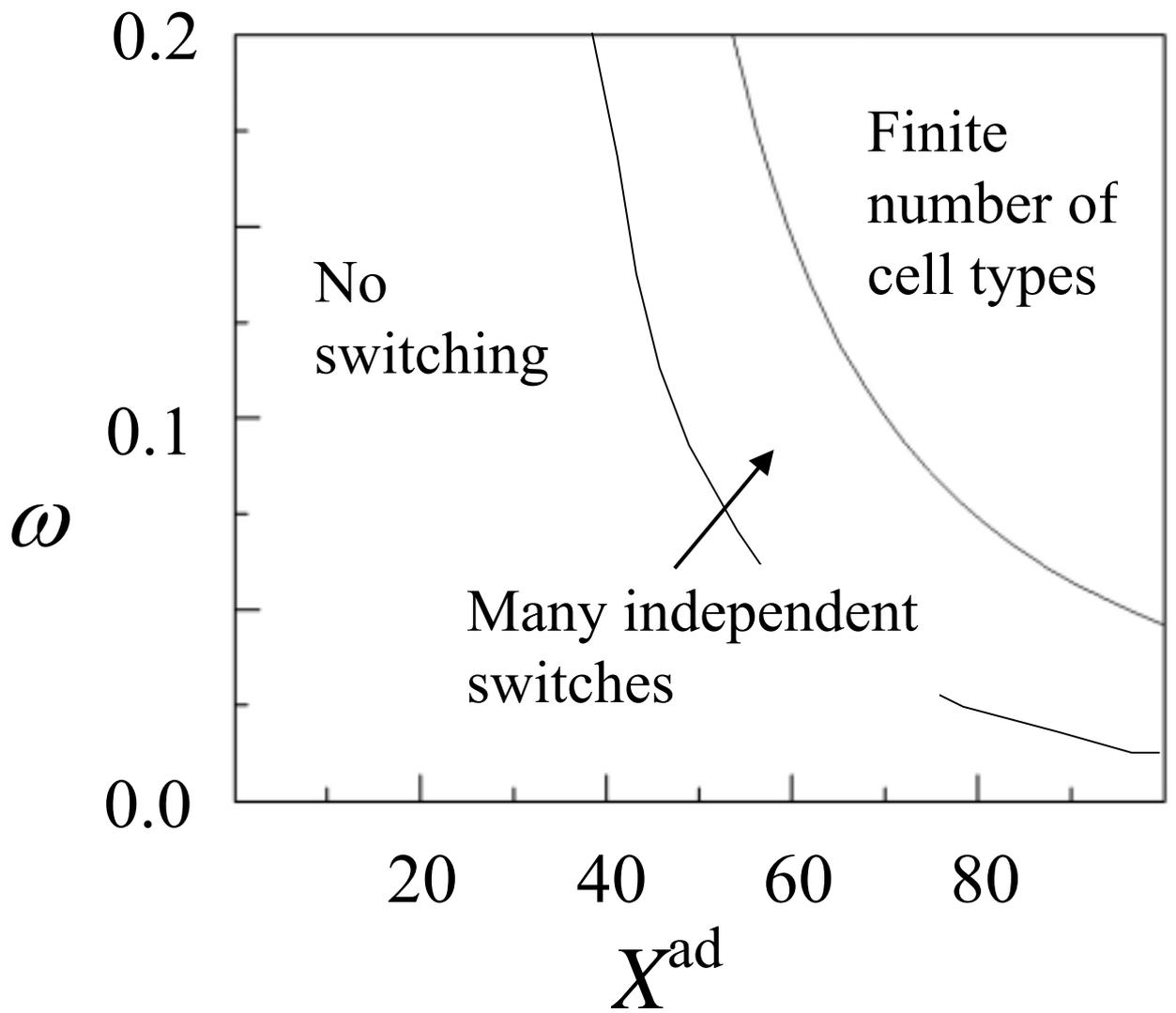

Fig.3  Sasai and Wolynes